\documentclass[10pt, preprint]{sigplanconf}

\usepackage{amsmath}
\usepackage{graphicx}
\newcommand{\REM}[1]{}

\begin{document}



\title{SMAT: An Input Adaptive Sparse Matrix-Vector Multiplication Auto-Tuner}

\authorinfo{Jiajia Li, Xiuxia Zhang, Guangming Tan, Mingyu Chen}
           {Institute of Computing Technology, Chinese Academy of Sciences}
           {lijiajia@ict.ac.cn}

\maketitle

\begin{abstract}
Sparse matrix vector multiplication (SpMV) is an important kernel in scientific and engineering applications. The previous optimizations are sparse matrix format specific and expose the choice of the best format to application programmers. In this work we develop an auto-tuning framework to bridge gap between the specific optimized kernels and their general-purpose use. We propose an SpMV auto-tuner (SMAT) that provides an unified interface based on compressed sparse row (CSR) to programmers by implicitly choosing the best format and the fastest implementation of any input sparse matrix in runtime. SMAT leverage a data mining model, which is formulated based on a set of performance parameters extracted from 2373 matrices in UF sparse matrix collection, to fast search the best combination. The experiments show that SMAT achieves the maximum performance of 75 GFLOP/s in single-precision and 33 GFLOP/s in double-precision on Intel, and 41 GFLOP/s in single-precision and 34 GFLOP/s in double-precision on AMD. Compared with the sparse functions in MKL library, SMAT runs faster by more than 3 times.
\end{abstract}




\section{Introduction}
\label{sec:intro}

Sparse Matrix Vector Multiplication (``SpMV") is one of the most important kernels in scientific and engineering areas. In lots of applications, SpMV plays an important role in their overall performance. For example, SpMV performance is a critical factor in electromagnetic field computation, laser fusion, fluid dynamics, climate simulation, and so forth. As a critical component of these applications, numerical solvers are the main focus in this paper. Numerical solvers also depend on SpMV performance due to a large percentage of execution time SpMV consumes. Take algebraic multi-grid algorithm for example, which is a iterator algorithm widely used in the above applications, the time percentage of SpMV is above 90\% of the overall algorithm. Therefore, it is meaningful to optimize SpMV, and make numerical solvers become application-aware and architecture-aware by automatically calling the best SpMV kernel.

Plenty of work has been dedicated to optimizing SpMV performance since 1970s, including the improvements of storage formats~\cite{OSKI, Vuduc_2005, CSX, Choi_2010, clspmv, Buluc_2011}, which is essential to SpMV behavior, and the optimizations considering novel computer architectures~\cite{Williams_2009, Bell:SpMV:NVIDIA:2008, Krishna_2011}. However, SpMV optimizations are rarely found applied widespread in numerical solvers. According to our knowledge, there is a performance gap between optimized SpMV kernels in literature and SpMV kernels used in numerical solvers. That is, the SpMV kernel called in solvers cannot achieve such good performance as the optimized SpMV in literature. The SpMV kernel in hypre library of LLNL is naively implemented in CSR format. Its performance is poor compared with the optimized SpMV kernel in~\cite{sparsity}, and the performance gap can reach to nearly 2 times.

There are several reasons for this phenomenon that current numerical solvers have not used the best optimized SpMV kernels. First and foremost, although SpMV optimization methods improve its performance, however, most of them focus on a specified storage format or sparse matrices in a similar type. Choi et al.~\cite{Choi_2010} improved matrix performance by improving SBELL instead of ELL format, while Eun-jin Im et al. focused on matrices with many dense blocks~\cite{sparsity}. The provisos prevent the corresponding optimization methods from being applied to numerical solvers with different application callers. Second, there are sometimes more than one type of sparse matrices used and generated in a realistic application. Take algebraic multi-grid (AMG) algorithm as an example, it generates a series of sparse matrices on several grid levels, and use them in the following calculations. The diverse matrices generated by the coarsen algorithm of AMG makes a single optimized SpMV kernel lose its effectiveness for the series of sparse matrices on different levels. Therefore, it is necessary to utilize more than one optimization method to improve the whole application performance. Finally, not only the storage formats and optimization methods influence SpMV performance, but different platforms also play a critical part in it. The same optimization and even implementation may behave differently in diverse platforms. Although this point has gained well studies and several successful auto-tuning libraries (i.e. SPARSITY~\cite{sparsity}, OSKI~\cite{OSKI}, clSpMV~\cite{clspmv}, etc.) are available, the premise is that users statically specify the sparse matrix format. With respect to the fact that sparse matrix format is application specific and may change dynamically, it is a must to choose optimization strategies dynamically and automatically for the best performance when being applied in numerical solvers and realistic applications on different platforms.

In this paper, we develop an SpMV Auto-Tuner (SMAT) to automatically choose the ``best" storage format and the ``best" SpMV implementation on X86 platforms. In this way, SMAT is able to provide the most suitable SpMV kernel for a numerical solver, according to its ability of application-awareness and architecture-awareness. Specifically, we make the following three main contributions in this paper.
\begin{itemize}
  \item We propose an input adaptive SpMV Auto-Tuner (SMAT) framework that is able to choose the ``best" format and implementation of SpMV. SMAT provides an unified interface based on CSR format and liberates users from choosing the best storage format.
  \item We extract a set of parameters to represent SpMV's performance characteristics from 2373 matrices in UF sparse matrix collection. These parameters are used to formulate the selection of the best SpMV kernel to a data mining model, which is implemented in our SMAT system.
    \item SMAT achieves the maximum performance of 75 GFLOP/s in single-precision and 33 GFLOP/s in double-precision on Intel, and 41 GFLOP/s in single-precision and 34 GFLOP/s in double-precision on AMD. Compared with the sparse functions in MKL library, SMAT runs faster by more than 3 times. The overhead of runtime auto-tuning can be amortized when the number of iterations is more than 29 in realistic applications.
\end{itemize}
\vspace{-0.25cm}

The rest of the paper is organized as follows. Section~\ref{sec:backgroud} presents the storage formats of sparse matrix, and the reason we choose them. Besides, UF sparse matrix collection is also introduced in this section. In Section~\ref{sec:SMAT} we propose the overview of SMAT system. We first analyze the sparse matrix characteristics and extract a set of representative parameters of the four storage formats in Section~\ref{sec:observations}. The details of SMAT system are explained in Section~\ref{sec:decision_tree} and Section~\ref{sec:runtime}, which mainly illustrate the data mining process of offline stage and the runtime process respectively. Section~\ref{sec:exp} illustrates the performance results and analysis. Related work is presented in Section~\ref{sec:related_work}, and conclusion in Section~\ref{sec:conclusion}. 

\vspace{-0.25cm}
\section{Background and Motivation}
\label{sec:backgroud}

\subsection{Storage Formats}

\begin{figure}[htbp]
\centering
  \includegraphics[scale=0.75]{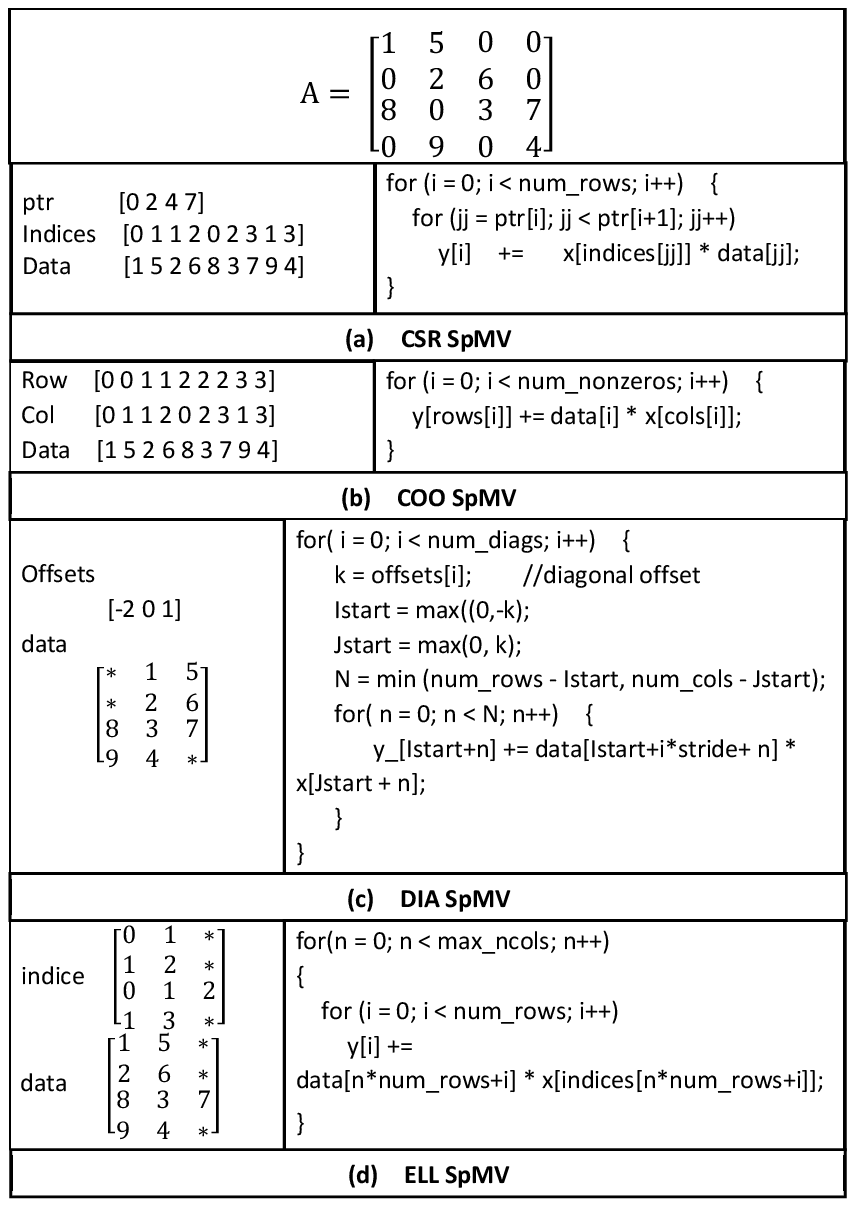}
  \caption{\small {Data structures of the four storage formats and their corresponding SpMV implementations}}
  \label{fig:kernels}
\end{figure}

To reduce complexity of space and computation, sparse matrices are always stored in a compressed way, where only nonzero elements are stored in a compressed data structure. Although tens of formats are developed since 1970s to date, four basic storage formats are extensively used: {\tt CSR, DIA, ELL, COO}. According to specific matrix features, some variants are derived from these basic formats.
Blocking storage formats are generated from the basic formats, such as BCSR is blocking CSR format and BDIA is blocking DIA format. Some formats need a sparse matrix to be reordered (JAD~\cite{sparskit}, CSX~\cite{Krishna_2011}) or divided (PTK, HYB~\cite{Bell:SpMV:NVIDIA:2008}, Cocktail~\cite{clspmv}), and then use existed formats after reordering or dividing. Besides, most of numeric solver packages are developed with the four basic storage formats, like MKL, wherein CSR is the most widely used. Therefore, we only consider the four basic storage formats in this paper, and make it possible to extend other formats in the future. The compressed data structure of the four formats is given in Figure~\ref{fig:kernels} accompanied with the corresponding SpMV implementation.

\begin{table}[htbp]
\centering
\caption{\small {Application distribution of UF collection}}
\label{tab:apps}
{\scriptsize
\begin{tabular}{|c|c|}
\hline
\textbf{Applications} & \textbf{number of matrices} \\\hline
linear programming & 327 \\\hline
graph & 323 \\\hline
structural & 277 \\\hline
combinatorial & 266 \\\hline
circuit simulation & 260 \\\hline
computational fluid dynamics & 168 \\\hline
optimization & 138 \\\hline
2D\_3D & 121 \\\hline
economic & 71 \\\hline
model reduction & 70 \\\hline
chemical process simulation & 64 \\\hline
power network & 61 \\\hline
theoretical quantum chemistry & 47 \\\hline
electromagnetics & 33 \\\hline
semiconductor device & 33 \\\hline
thermal & 29 \\\hline
materials & 26 \\\hline
least squares & 21 \\\hline
computer graphics vision & 12 \\\hline
statistical mathematical & 10 \\\hline
counter-example & 8 \\\hline
acoustics & 7 \\\hline
biochemical network & 3 \\\hline
robotics & 3 \\\hline
\end{tabular}
}
\end{table}
\subsection{UF collection}

UF collection collects a suit of sparse matrices extracted from realistic applications since 1991. It is built to bridge the gap between computational scientists and sparse matrix algorithm developers. Compared with other sparse matrix collections, such as Matrix Market~\cite{matrixmarket} and Harwell-Boeing Collection~\cite{Harwell-Boeing}, UF collection has a larger matrix size, and covers more application areas. For simplicity without loss of accuracy, we exclude the matrices with complex values and too small size. Totally 2373 sparse matrices are studied in our work. Table~\ref{tab:apps} lists their application areas, which cover more than 20 application domains in scientific and engineering.

\subsection{Motivation}
\begin{figure}[htbp]
\centering
  \includegraphics[scale=0.4]{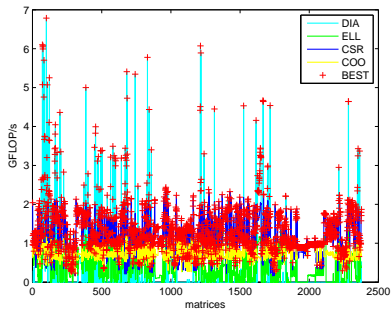}
  \caption{\small {Performance of the four basic storage formats.}}
  \label{fig:perf_comparision}
\end{figure}

A storage format has its own application scope among sparse matrices with diverse characteristics, and achieves the best SpMV performance in a subset of it. We setup an experiment to present a quantitative performance difference among the four storage formats. Figure~\ref{fig:perf_comparision} plots the performance in GFLOP/s for the four storage formats applied to the 2373 matrices. The performance difference indicates that it is not fair to provide one storage format in sparse numeric solvers.

\vspace{-0.3cm}

In fact, both DIA and ELL formats need zero filling operations so that some matrices are not suitable to be represented by them. According to the program of ~\cite{Bell:SpMV:NVIDIA:2008}, the filling ratio of zeros satisfies that the number of diagonals (DIA) or the maximum number of nonzeros per row (ELL) should not be larger than 20 times of the average number of nonzeros per row. This ensures the ratio of nonzeros in their data structures will not lower than 5\%, and we consider this limitation is practicable. Thus, with respect to the filling ratio, only part of the 2373 matrices are suitable to either DIA or ELL formats. Since both CSR and COO only store nonzero elements of a sparse matrix, they are suitable for all sparse matrices. Note that the ``suitable" format does not mean the ``best" one. Table~\ref{tab:subsets} summarizes the difference. To some extent, both DIA and ELL are special storage formats compared with CSR and COO, then intuitively the usage of them is expected to achieve better performance. On the contrast, we observe that a small portion of them shows performance advantage. For example, only 218 of 458 DIA matrices and 299 of 1878 ELL matrices, achieve the best performance. Similar situations are observed for the cases of CSR and COO.

\begin{table}[htbp]
\centering
\caption{\small {Matrix classification of the four formats}}
\label{tab:subsets}
{\tiny
\begin{tabular}{|c|c|c|c|c|}
\hline
{\bf Storage Formats} & DIA & ELL & CSR & COO \\\hline
{\bf \scriptsize suitable} & 458 & 1878 & 2373 & 2373 \\
 & (DIA\_mats) & (ELL\_mats) & (all\_mats) & (all\_mats) \\\hline
{\bf \scriptsize best}  & 218 & 299 & 1458 & 603 \\
& (good\_DIA\_mats) & (good\_ELL\_mats) & (good\_CSR\_mats) & (good\_COO\_mats) \\\hline
\end{tabular}
}
\end{table}

Therefore, the first question is to {\em determine its best storage format for any given sparse matrix}. Based on a selected format, there are already well-studied auto-tuning tools or libraries can be leveraged to obtain its best implementation on a specific platform. If the matrix format keep the same during the lifetime of an application, even a brute-force search algorithm is worth to be applied to get the best one. As noted before, some solver algorithms like AMG change the distribution of non-zeros of matrices dynamically. Therefore, the second question is to {\em search the best combination of the best format and implementation in runtime with low overhead}.

\section{Overview of SMAT System}
\label{sec:SMAT}
\begin{figure}[htbp]
\centering
  \includegraphics[scale=0.35]{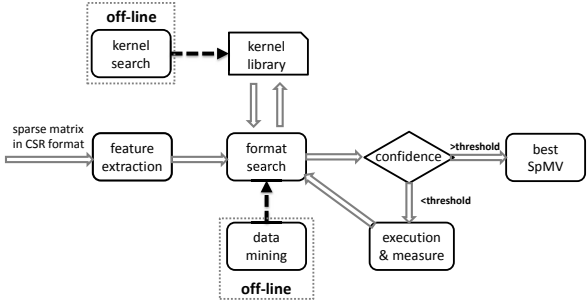}
  \caption{\small {The framework of SMAT system.}}
  \label{fig:smat_new}
\end{figure}

With respect to the fact that most of numeric solvers implement CSR as their fundamental storage format of sparse matrices, our SMAT framework specifies an unified interface with CSR format. Application programmers do not need to care about the choice among the formats, but prepare the input matrices in CSR format. The SMAT auto-tuner is in charge of selecting the best format and implementation. Figure~\ref{fig:smat_new} depicts a high level framework of the SMAT system. SMAT uses a hybrid auto-tuning strategy of static and dynamic optimizations. Correspondingly, it is composed of two separated stages: off-line and runtime.

The off-line functions are performed when the auto-tuning system is installed. The kernel search component can make full use of the well-studied auto-tuning tools to generate a high performance kernel library on a specific architecture. The kernel library holds a symbol table that maps each storage format to its best code implementation. Once the best format is selected, its best implementation can be found in the kernel library. Another component in the off-line stage is to build a model used to search the best format in runtime. We leverage data mining approaches to generate a decision tree for format search unit in runtime. The training matrices are chosen from UF collection. In runtime stage, we use the generated decision tree to estimate the best storage format among the decided SpMV implementations. With regard to the possible inaccurate predictions, a refinement of the execute and measure method of auto-tuning is triggered.

A key premise of the SMAT system is a set of appropriate parameters that is used to build a data mining model. These parameters should represent comprehensive characteristics of a sparse matrix, and are closely related with its SpMV performance. We extract the parameters from the four storage formats by measuring the performance variation of the SpMV implementations. In the following context, we first describe how to extract the performance parameters.

\section{Parameters Extraction}
\label{sec:observations}

Table~\ref{tab:parameters} summarizes the final set of parameters extracted for the SMAT system. First we define notations for several basic parameters: M (number of rows), N (number of columns), NNZ (number of nonzeros) and aver\_RD (average number of nonzeros per row). Some of which will be used to derive other feature parameters. The table lists 10 feature parameters and their applicability in each storage format. For example, the parameter set of $\{M,NNZ,Ndiags,NTdiags\_ratio,ER\_DIA\}$ acts as a key criterion in the decision tree to measure the characteristics of DIA format. Since the default format is CSR in the SMAT system, the parameter extraction is conducted on other three formats of DIA, ELL and COO.

\begin{table}[htbp]
\centering
{\scriptsize
\caption{\small {Feature parameters of a sparse matrix and the relationship with the formats}}
\label{tab:parameters}
\begin{tabular}{|c|p{70pt}|c|c|c|c|}
\hline
\textbf{Parameter} & \textbf{Meaning} & \textbf{DIA} & \textbf{ELL} & \textbf{CSR} & \textbf{COO} \\\hline
M & the number of rows & $\surd$ & $\surd$ & $\surd$ & $\surd$ \\\hline
NNZ & the number of nonzeros & $\surd$ & $\surd$ & $\surd$ & $\surd$ \\\hline
Ndiags & the number of diagonals & $\surd$ & & & \\\hline
NTdiags\_ratio & the ratio of ``true" diagonals to total diagonals  & $\surd$ & & & \\\hline
ER\_DIA & the ratio of nonzeros in DIA data structure  & $\surd$ & & & \\\hline
max\_RD & the maximum number of nonzeros per row  & & $\surd$ & & \\\hline
min\_RD & the minimum number of nonzeros per row & & $\surd$ & & \\\hline
var\_RD & the variation of the number of nonzeros per row  & & $\surd$ & & \\\hline
ER\_ELL & the ratio of nonzeros in ELL data structure  & & $\surd$ & & \\\hline
R & a factor of power-law distribution  & & & & $\surd$ \\\hline
\end{tabular}
}
\vspace{-0.6cm}
\end{table}

\begin{table}[htbp]
\centering
\caption{\small {Performance features of the four formats}}
\label{tab:kernel_comparison}
{\scriptsize
\begin{tabular}{|p{50pt}|p{70pt}|p{80pt}|}
\hline
{\bf storage formats} & {\bf extra computation} & {\bf repeated times of writing Y} \\\hline
DIA &  decided by zero-filling & number of diagonals \\\hline
ELL &  decided by zero-filling & maximum number of nonzeros per row \\\hline
CSR &  no & 1 \\\hline
COO &  no & indirect \\\hline
\end{tabular}
}
\vspace{-0.25cm}
\end{table}

\begin{figure}[htbp]
\begin{tabular}{c|c}
  \includegraphics[scale=0.26]{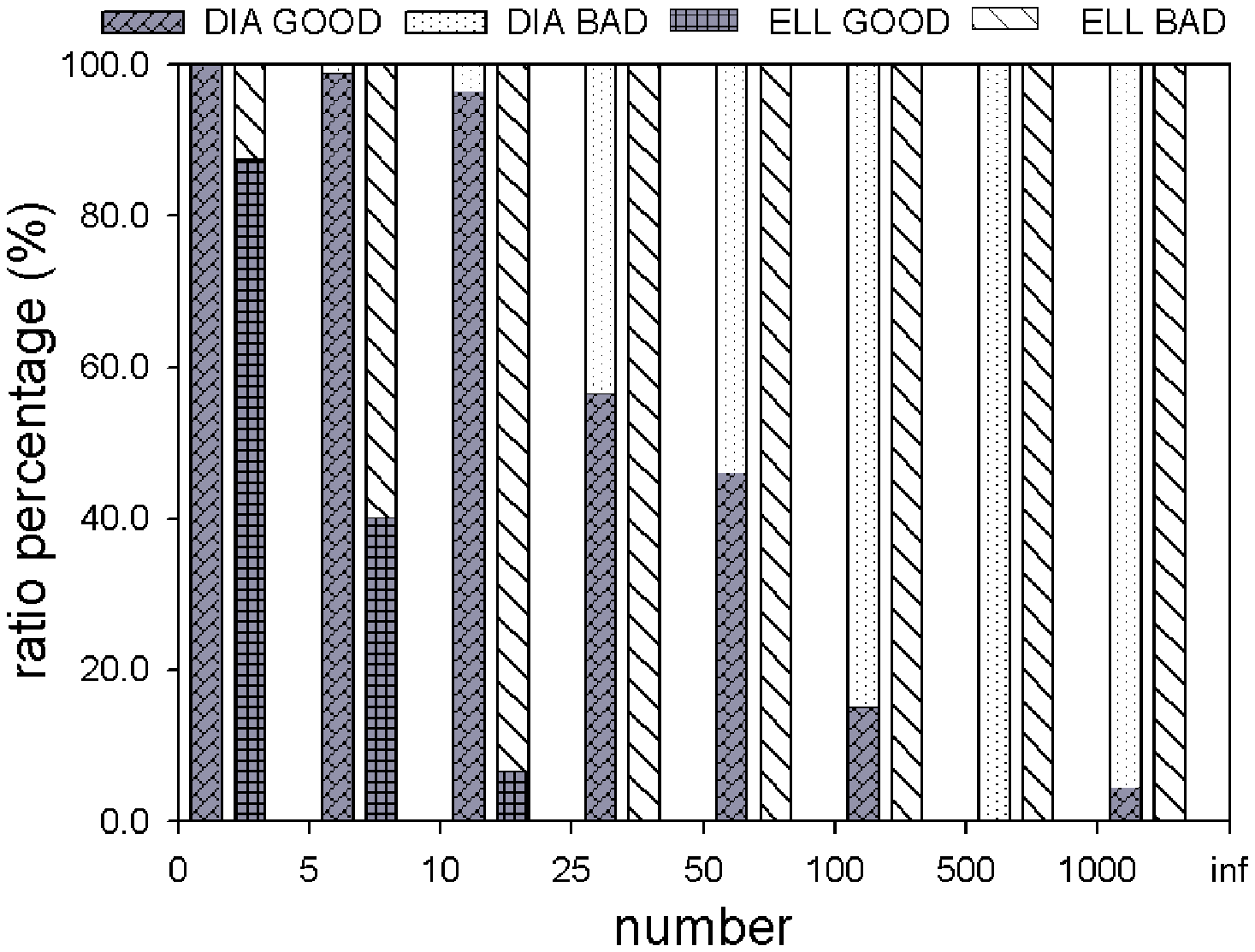}
&
  \includegraphics[scale=0.25]{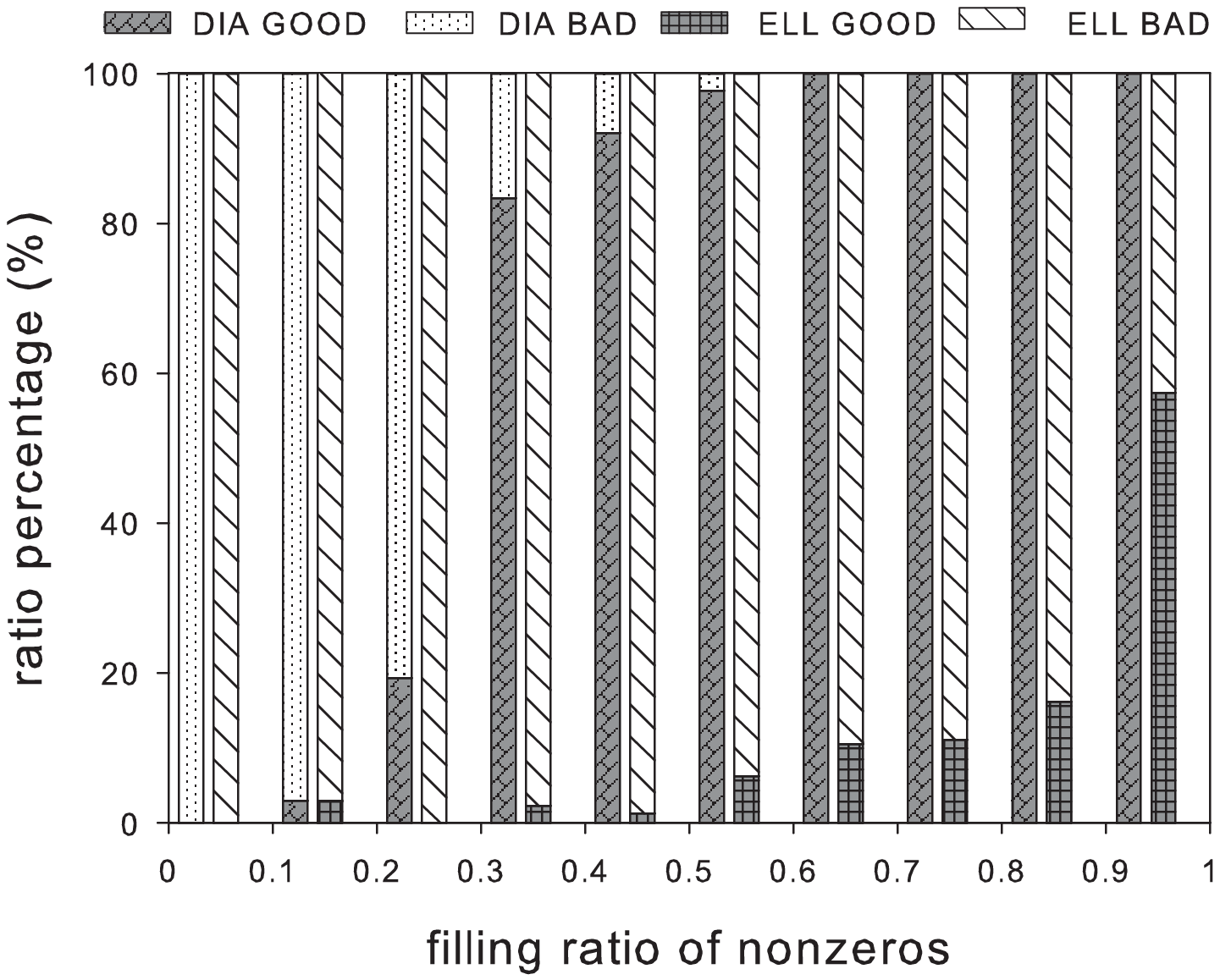}
\\
(a) \small {Ndiags and max\_RD} & (b) \small {ER\_DIA and ER\_ELL}
  \end{tabular}
  \caption{\small {The influence of Ndiags and ER\_DIA on DIA-SpMV and max\_RD and ER\_ELL on ELL-SpMV}}
  \label{fig:four_paras}
  \vspace{-0.2cm}
\end{figure}

According to the rule of zeros filling ratio measured in ~\cite{Bell:SpMV:NVIDIA:2008} and our experimental results in Table~\ref{tab:subsets}, the matrices, whose maximum number of diagonals is larger than $20 \times aver\_RD$, are excluded from both DIA and ELL candidates. By investigating the SpMV kernels in Figure~\ref{fig:kernels}, the advantage of both DIA and ELL comes from the regular data accesses to the matrix and X-vector. However, as summarized in Table~\ref{tab:kernel_comparison} it incurs extra computations by filling zeros and extra memory operations on Y-vector by accumulating Y-vector in the outer-loop. Therefore, for DIA matrices, the number of duplicated writing of Y-vector is determined by the number of diagonals {\em Ndiags}, and the extra computations are determined by the ratio of nonzeros in a matrix {\em ER\_DIA}=$\frac{NNZ}{Ndiags\times M}$. For ELL matrices, the number of duplicated writing of Y-vector is determined by the maximum number of nonzeros per row {\em max\_RD}=$\max_{1}^{M}\{number~of~nonzeros~per~row\}$, and the extra computations are determined by the ratio of nonzeros in a matrix {\em ER\_ELL} = $\frac{NNZ}{(max\_RD \times M)}$.

\REM{
\begin{figure}[htbp]
\centering
  \includegraphics[scale=0.35]{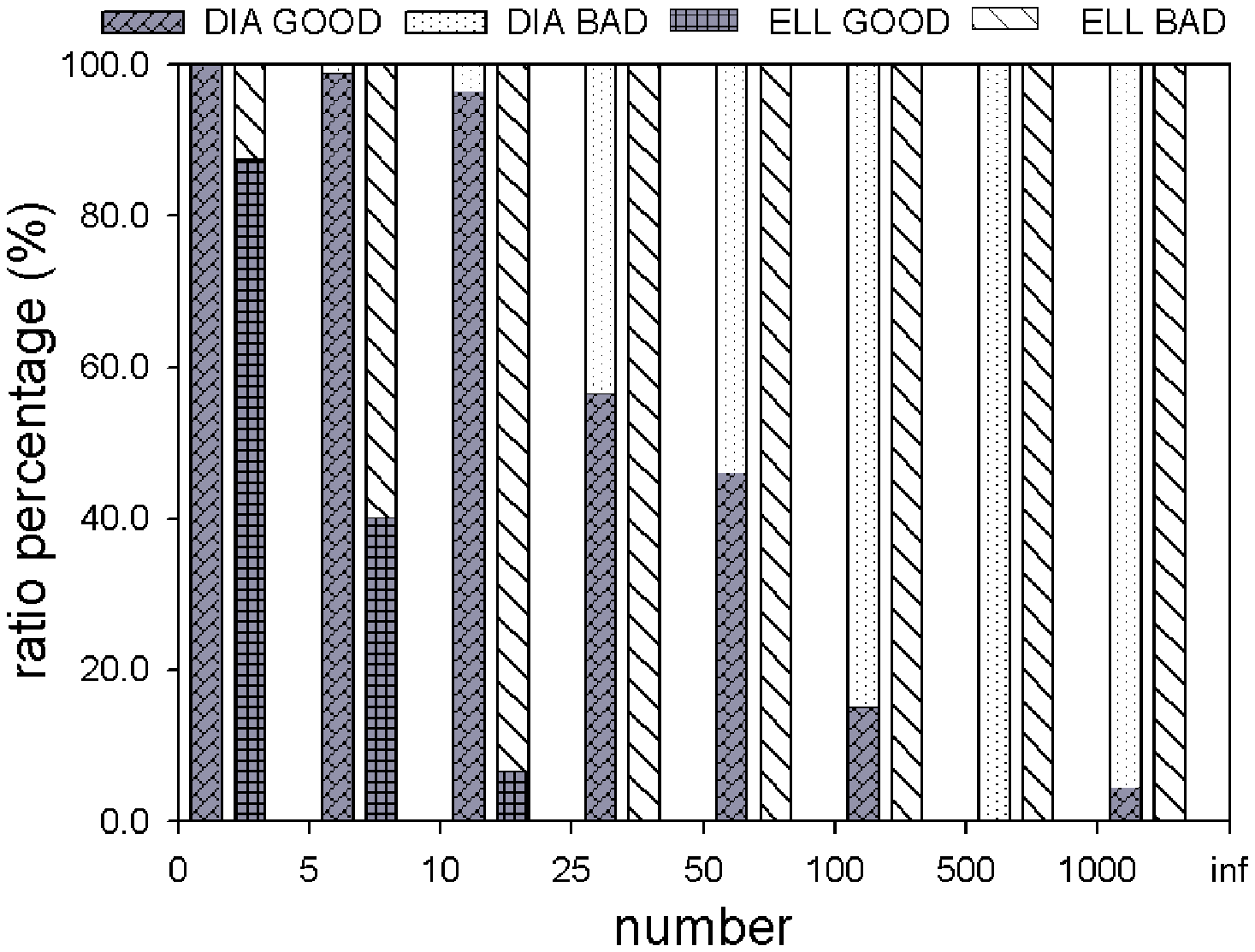}
  \caption{\small The influence of Ndiags on DIA-SpMV and max\_RD on ELL-SpMV}
  \label{fig:Ndiags_max_RD}
\end{figure}
}

The strategy to estimate values of the extracted parameters is to perform statistical analysis on experimental results from testing the 2373 sparse matrices. Figure~\ref{fig:four_paras} plots the effects of different configurations of parameters. In these figures, ``GOOD" bars mean the matrices which achieve better performance in either DIA-SpMV or ELL-SpMV than other formats, and they belong to ``good\_DIA\_mats" or ``good\_ELL\_mats" in Table~\ref{tab:subsets}. Otherwise, the matrix is tagged by ``BAD" that means they fall into candidates of either CSR or COO. Based on the extensive experiments, we observe that:
\begin{itemize}
   \item When $Ndiags<25$, nearly all of matrices will benefit from DIA format. When $Ndiags$ is larger than 500, few matrices benefit from it.
   \item When $ER\_DIA>60\%$, DIA format totally wins. If the ratio is too small, there is no advantage to use DIA format.
   \item When $max\_RD<5$, it is most possible for ELL-SpMV to achieve best performance. Otherwise, ELL-SpMV rarely shows its advantage.
    \item ELL may show its benefits only if $ER\_ELL>90\%$.
 \end{itemize}

\begin{figure}[htbp]
\begin{tabular}{c|c}
  \includegraphics[scale=0.25]{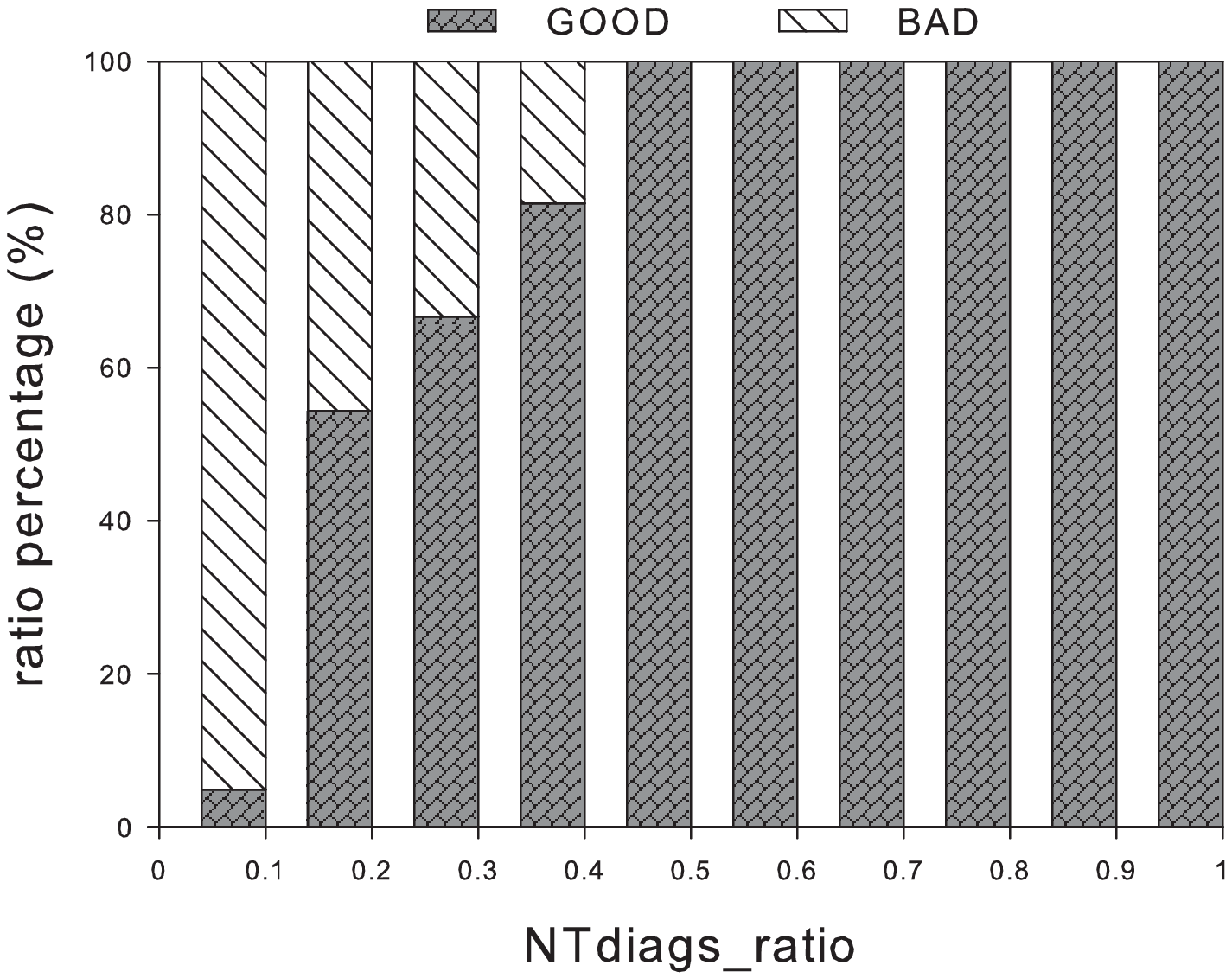}
&
  \includegraphics[scale=0.25]{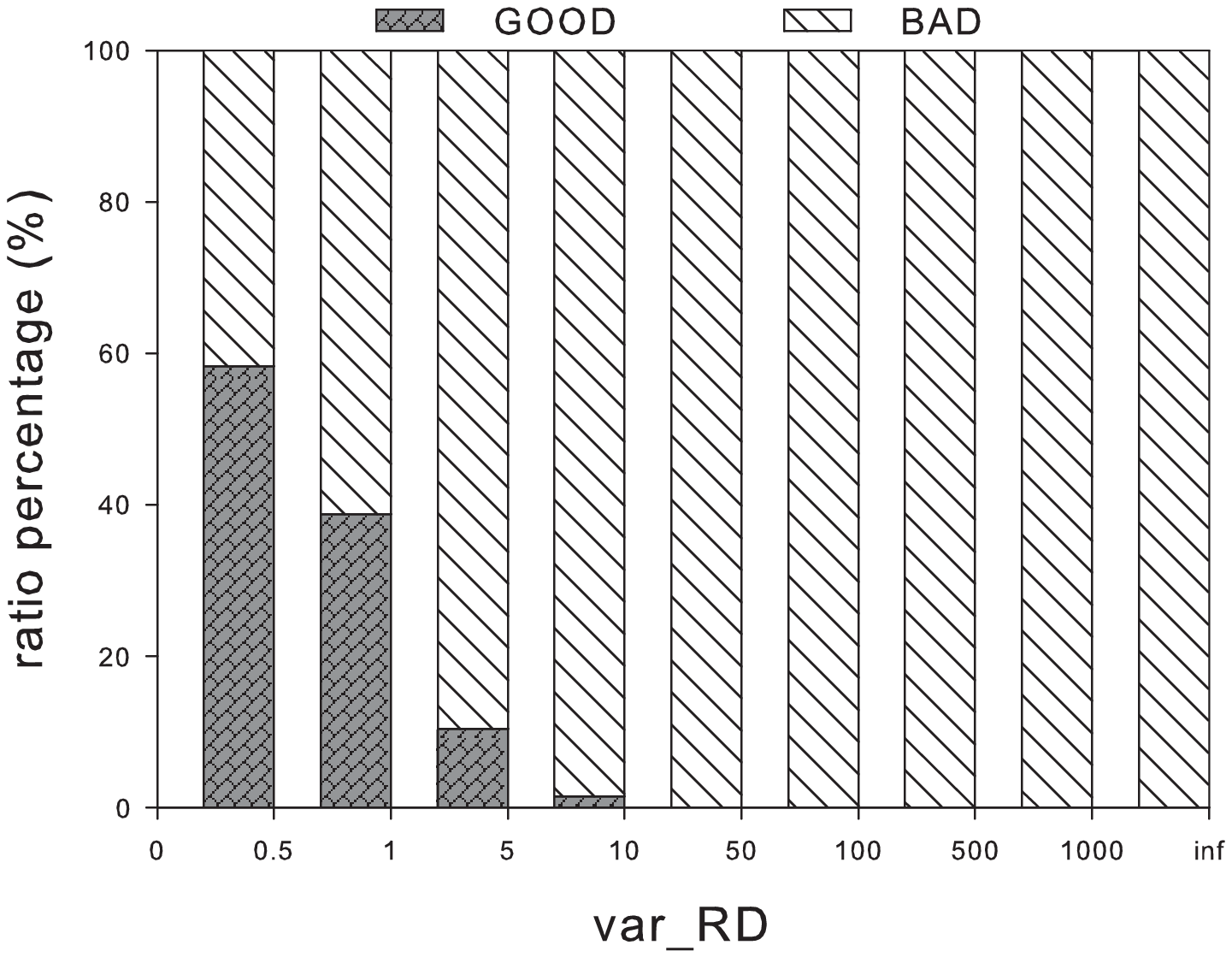}
\\
(a) \small {NTdiags\_ratio} & (b) \small {var\_RD}
  \end{tabular}
  \caption{\small {The influence of NTdiags\_ratio on DIA-SpMV and var\_RD on ELL-SpMV}}
  \label{fig:two_paras}
\end{figure}

Undoubtedly, only using these initial parameters may result in miss predication, more parameters need to be extracted for a high accuracy. Looking at Figure~\ref{fig:four_paras}, we observe that the performance of DIA-SpMV still beats SpMV in other formats although the {\em ER\_DIA} value is not large. Take Bai/ck400 as an example, the ER\_DIA is only 26\% while its performance in DIA format is 1.5 times of that in CSR format. Therefore, a new definition and parameter are introduced -- ``true diagonal" and {\em NTdiags\_ratio}. ``true diagonal" means a diagonal the nonzero ratio of which is larger than a threshold. When DIA-SpMV implemented, this diagonal will achieve much higher performance than others. {\em NTdiags\_ratio}=$\frac{number~of~``true~diagonals"}{Ndiags}$, represents the ratio of the number of ``true diagonals" to the number of total diagonals.
Though both {\em NTdiags\_ratio} and {\em ER\_DIA} have relationships to the nonzero ratio of DIA data structure, {\em NTdiags\_ratio} helps to estimate the behavior of DIA-SpMV better and more accurate. The influence of {\em NTdiags\_ratio} is shown in Figure~\ref{fig:two_paras}(a). We observe that:
\begin{itemize}
  \item When $NTdiags\_ratio>40\%$, SpMV benefits by using DIA format.
\end{itemize}

Similar to DIA, we extract another parameter for ELL format. Considering the characteristics of ELL format, which is suitable to a matrix with similar number of nonzeros per row, it is needed to introduce the variation of nonzero numbers per row {\em var\_RD} = $\frac{\Sigma_1^M|row\_degree - aver\_RD|^2}{M }$. The influence of var\_RD on ELL-SpMV performance is shown in Figure~\ref{fig:two_paras}(b). We observe that:
\begin{itemize}
  \item Only if $var\_RD<0.5$, ELL format will show its benefit.
\end{itemize}

Until now we filter DIA and ELL matrices, the remaining format is COO. From~\cite{Yang_2011} we realize that COO format will achieve the highest performance among the four formats on NVIDIA GPU, when the input sparse matrix is a representation of small-world network. Since the node degree of small-world network obeys power-law distribution, we use power-law distribution $P(k) \sim k^{-R}$ to decide whether the matrix is a small world network matrix. We calculate the R value through least squares method according to this equation and observe its value. If R value is in [1,4) (see~\cite{Aiello00arandom, Han_2005} for details), we consider this sparse matrix is a small-world network matrix.

So far, we extract a set of parameters to represent performance characteristics of a sparse matrix. The parameters have already been listed in Table~\ref{tab:parameters}. Although there are some obvious rules for part of the parameters, like the parameters of DIA and ELL, their relationship is complex. The more complicated and accurate rules will be dug out in the following section with the help of data mining method. 

\section{Decision Tree}
\label{sec:decision_tree}

In Section~\ref{sec:observations}, we extract a set of parameters to represent the features of a sparse matrix and obtain some observations from the performance measurement. Although through the observations we can obtain some useful rules (especially DIA and ELL formats) to guide format searching, there still exist formats that cannot be extracted parameters related to its SpMV performance. Therefore, it is necessary to utilize a data mining method to mine more detailed rules to better predict the best format. Besides, to generate a rule the threshold is requisite for each parameter, data mining method is able to get more accurate threshold value and more labor-saving than hand-tuning through lots of experiments. Based on our verification that choosing the best format is the classification problem in data mining area, we introduce data mining method to the SMAT system as a part of the offline stage. Through the data mining procedure, a decision tree is generated for on-line format searching process.

\subsection{Classification Problem}

The set of parameters in Table~\ref{tab:parameters} can be considered as a collection of attributes in data mining field. The values of the parameters are generated for each sparse matrix. From Table~\ref{tab:parameters}, each attribute represents a sole feature of a matrix, and each set of parameter values is a set of mutually exclusive classes. Therefore, the parameters and values constitute an attribute-value dataset, which is the input of a data mining tool.

Among the attributes, ``best\_format" attribute represents the best format in which the corresponding SpMV implementation achieves the highest performance. The possible values of this attribute are DIA, ELL, CSR, COO. Since it is the aim to choose the best storage format, the ``best\_format" is the target attribute, and we use it to make classifications. Now we need to find a mapping from the features of a given sparse matrix to the best format of it, and this mapping should be applied to the incoming sparse matrices with new features. With the expression of formulation, the mapping is described in Equation~\ref{equ:mapping}, where $\vec{x_i} (i=1, \ldots , n)$ represents a set of parameter values of a sparse matrix in the training set, and $\vec{TH}$ stands for the set of thresholds for each of the attribute. $C_n(DIA, ELL, CSR, COO)$ represents one categories of the four ones, which is our aim of classification.
\begin{equation}
\footnotesize
\label{equ:mapping}
    f( \vec{x_1}, \vec{x_2}, \ldots , \vec{x_n}, \vec{TH} ) \rightarrow C_n(DIA, ELL, CSR, COO)
\end{equation}

According to the above descriptions, choosing the best format is obviously a classification problem of data mining. We use training set to decide the values of thresholds $\vec{TH}$, and generate a decision tree. When a new sparse matrix comes, the features of it will be extracted and we are able to predict its category (the best format) using the decision tree.

\subsection{Decision tree generation}

\begin{figure}[htbp]
\centering
  \includegraphics[scale=0.25]{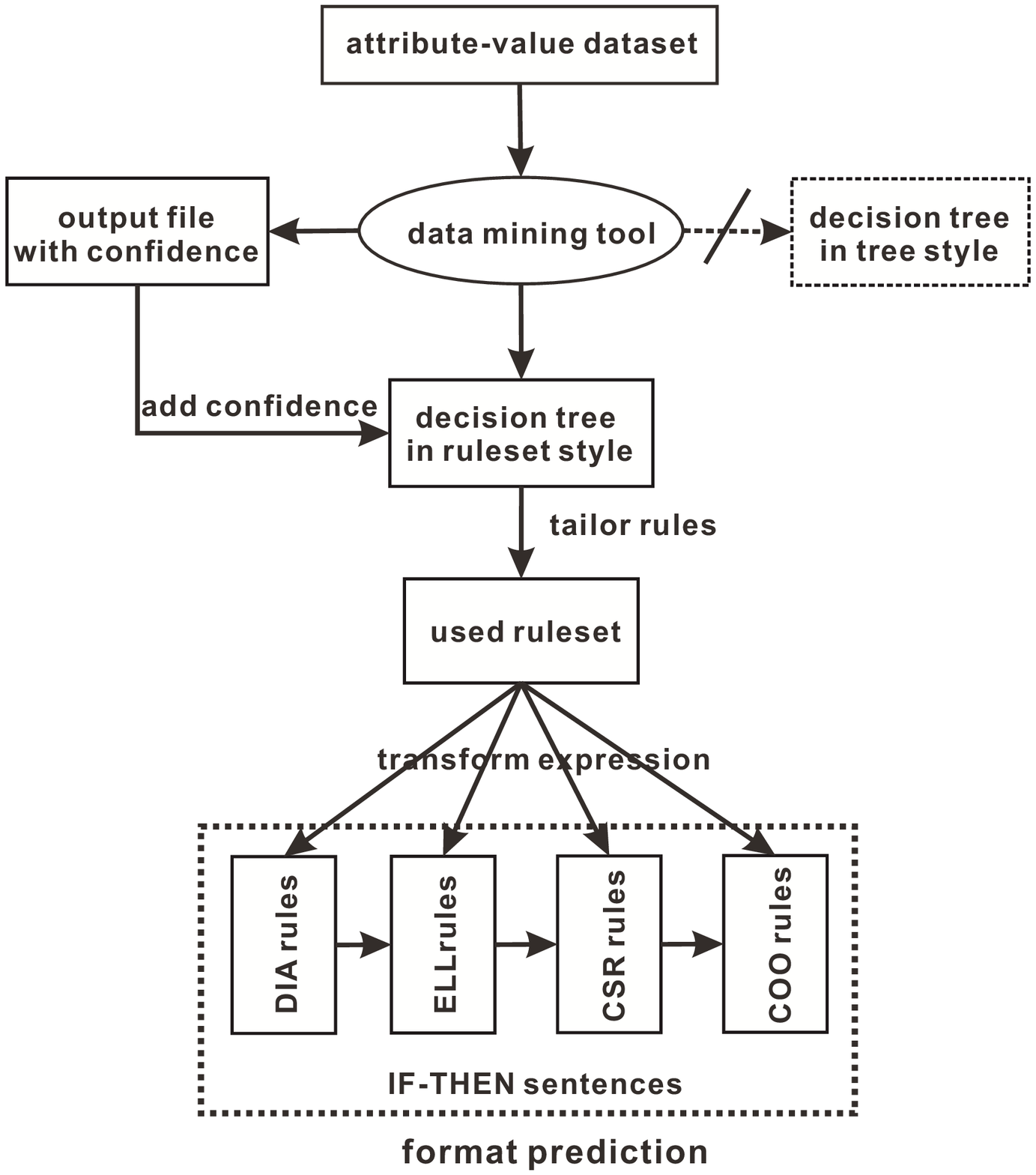}
  \caption{\small {The procedure of decision tree generation}}
  \label{fig:decision_tree}
\end{figure}

We use data mining tool C5.0~\cite{See_C5.0} to generate decision tree from the training set, which includes 2055 matrices of UF collection after 318 matrices excluded by proportion of each problem in it. The generated attribute-value dataset is the input of C5.0, and a decision tree is yielded by using the dataset. The procedure of the decision tree generating process is depicted in Figure~\ref{fig:decision_tree}, and we explain four main operations of this procedure in the next paragraphs.

\textit {Choosing Rulesets}~~~~From the input attribute-value dataset, C5.0 can generate two forms of decision tree. One is represented in a tree style, and the other is in ruleset style. Though the two styles represent similar information of the decision tree, ruleset has its advantages. First, ruleset classifiers are often more accurate than decision tree as illustrated in C5.0. Though it requires more computer time than generating a decision tree, it only affects the speed of offline stage which runs during the installment. Second, it is more convenient for us to integrate the rules into our code system, because it is much easier to convert the rules into IF-THEN sentences. So we choose ruleset to represent the generated decision tree in the SMAT system.

\textit {Adding Confidence}~~~~Since a rule may classify a matrix inaccurately, even in the training matrix set, the accuracy ratio of each rule is calculated and stored in the output file. The accuracy ratio, also called the confidence of a rule, is the ratio of the number of right classification matrices to the number of matrices which fall in this rule. The larger the confidence, the more accurate and believable of a rule. Moreover, a matrix may satisfy more than one rule at a time, and to decide a best rule the confidence value is used. C5.0 outputs two distinct files -- ruleset file and the output file. The ruleset file only assembles all the rules without the confidence of each rule. We update the ruleset by extracting the confidence from the output file.

\textit {Tailoring rules}~~~~As the generated ruleset may include several tens of rules, it is necessary to tailor the rules if part of them can achieve similar prediction accuracy. For example, in our test on Intel platform, Rule No.1 - No.15 make the error ratio decrease to 9.6\%, which is quite close to error ratio (9.0\%) achieved by all the 40 rules. So we order the rules by their estimated contribution to the overall training set. The rule that reduces the error rate most appears first and the rule that contributes least appears last. In this way, we extract the rules from top to down until the subset of rules achieve similar predictive accuracy (1\% accuracy gap is accepted) with all the rules. The selected rules are used in runtime stage of SMAT.

\textit {Transformation}~~~~We transform the rules to IF-THEN sentences which can be embedded in SMAT program. At the same time, the ruleset is divided into four subsets with each one belonging to a sole classification. In this way, though a matrix may fall into several rules with different confidence values, we can take the largest value as the the unique confidence of the chosen format. Thus, even a matrix is predicted suitable to different formats, the format confidence will discriminate the best one. Besides, according to the overall performance behavior and the predictability, we arrange the prediction procedure in the order of DIA, ELL, CSR, and COO (see the details in Section~\ref{sec:runtime}).

In summary, according to the input attribute-value dataset collected in the training matrix set, we finally obtain a set of IF-THEN sentences in a particular order, which is easily embedded into the runtime stage of SMAT. After the kernel search and data mining modules, the off-line stage is finished. In this stage, the platform diversity is reflected by the SpMV performance with different implementations and the ``best\_format" parameter value in the input dataset of the data mining tool. Note that the off-line stage only need to execute once when the SMAT system is transplanted to a new platform.

\REM{
\begin{itemize}
  \item \textit{Choosing Rulesets} Rulesets are chosen to represent the decision tree in C5.0. Since ruleset classifiers are often more accurate than decision trees, and more convenient to integrate the rules into our own codes system.~\cite{XX} So we choose ``-r" option to use in C5.0.
  \item \textit{Tailoring Rules} As the generated rulesets may include several tens of rules, and it is no need to use all of them to achieve the prediction accuracy on the same level. For example, in our test on Intel X5680, Rule No.1 - No.15 make the error ratio decrease to 9.6\%, which is quite close to error ratio (9.0\%) achieved by all the 40 rules. So we order the rules by estimated contribution to predictive accuracy using the ``-u" option. Under this option, the rule that most reduces the error rate appears first and the rule that contributes least appears last. In this way, we extract the front rules which achieve similar predictive accuracy (1\% accuracy gap is accepted) as the accuracy of all the rules. The selected rules are used in rumtine stage of SMAT.
  \item \textit{Adding Confidence Tag} Each rule has different prediction accuracy values and the differences are quite large for some time, so we attach a confidence value to each rule. If the confidence value of a rule is large enough, its classification is considered to be believable. Otherwise, we need other operations to make up for the inaccuracy phenomenon. This will be illustrated in detail in the next section.
\end{itemize}
}

\vspace{-0.25cm}
\section{Runtime Auto-Tuning}
\label{sec:runtime}

\begin{figure}[htbp]
\centering
  \includegraphics[scale=0.25]{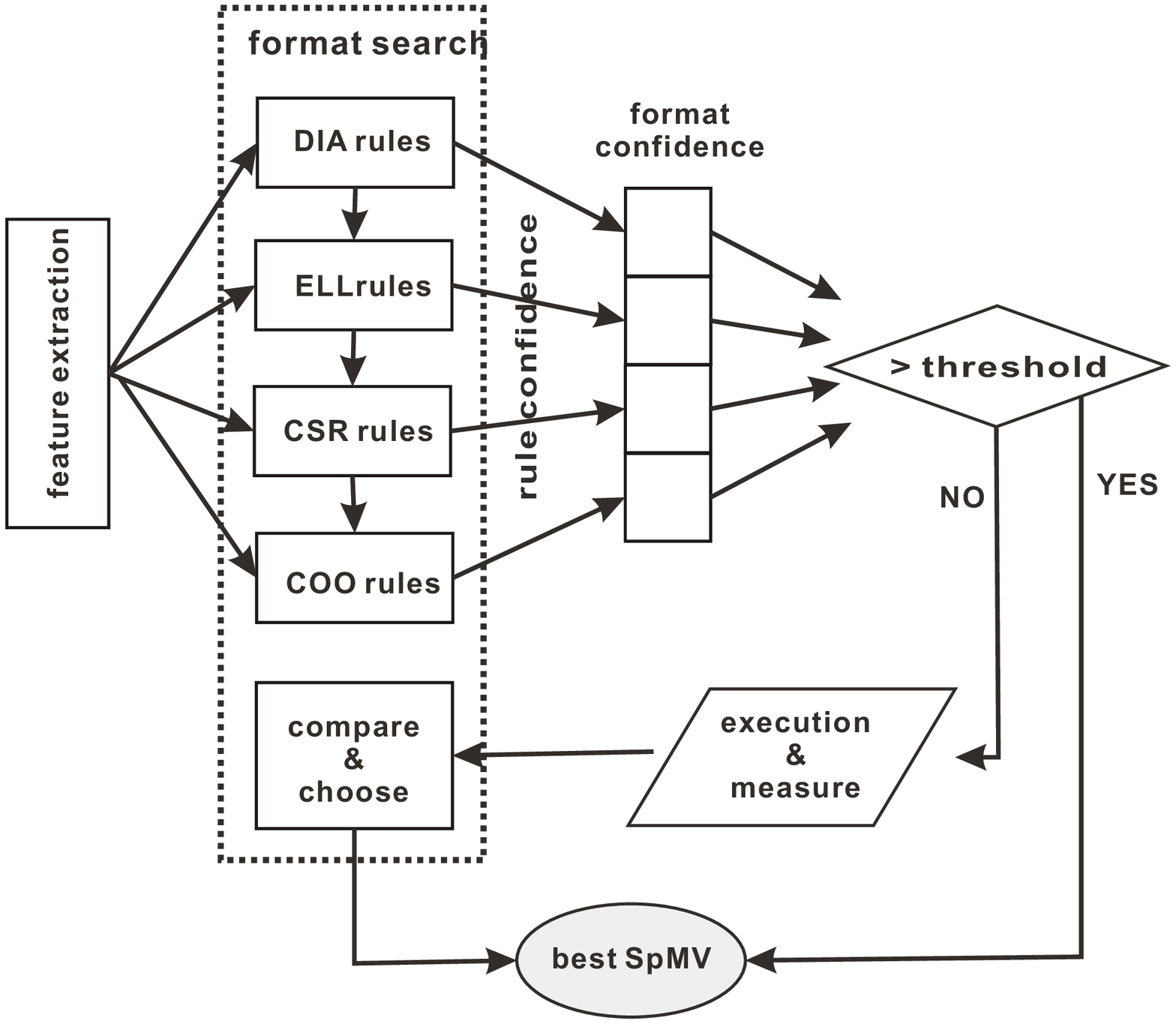}
  \caption{\small {Procedure of the runtime stage of SMAT}}
  \label{fig:runtime}
\end{figure}

As the main subject of the SMAT system, the runtime stage is shown in details in Figure~\ref{fig:runtime}. The input of it is a sparse matrix stored in CSR format, and the output is the best storage format and SpMV implementation of the input matrix. We explain the details in the following context.

During the runtime procedure, the features of the input sparse matrix are extracted in the feature extraction module. The extraction process is similar to the parameter value generating process in the offline stage. After the features are collected, SMAT proceeds the format search module which is generated from data mining process (Figure~\ref{fig:decision_tree}) in IF-THEN pattern. Due to the ralationship between confidence values of each format and the threshold value, two paths will be chosen. If the format confidence is larger than the threshold, we consider the best format is found, and output the format and its SpMV implementation. Otherwise, if the format confidence is smaller than the threshold, the execution\&measure module will be used to execute one or more SpMV implementations for once and measure its/their performance. The compare\&choose composition of format search module does the comparison among the performance numbers, and choose the format with the highest performance. That is the overall procedure of the runtime stage.

However, considered the time consuming of the feature extraction process, there is no need to wait until all the features of a sparse matrix be gathered, and then execute the format search module. Instead, a time-saving strategy is introduced, and only part of feature extraction and format search module need to be processed, once its format confidence is larger than the threshold. For example, when the feature extraction module extracts enough features for DIA format, the DIA rules of format search module is triggered to be active, and predicts if the input matrix satisfies the DIA ruleset. If format confidence (the largest value of all the rule confidence in DIA ruleset) is larger than the threshold, DIA format and the corresponding SpMV implementation are considered to be the ``best" SpMV, which will be the output of SMAT system.

Note that format search module not only collects the rules of all the four formats, but also includes the order of the four rulesets. To decide the format confidence, we divide the whole ruleset into four subsets corresponding to the four formats. The four subsets have two specificities. One is the independence, the other is the order. That is the subsets should be independent with each other and processed in order. As for independence, it is possible for each format to calculate its format confidence independently. Thus, the criterion is generated which we replied on to predict the best format. The order among the subsets makes the former format(s) calculating its format confidence as fast as possible. Therefore, there is a chance that the latter format(s) need not to do feature extraction and format perdiction any more, since the best format have been generated from the former formats. And the time of runtime stage is saved. The order of the four subsets are determined by their performance and prediction accuracy. As we know, DIA achieves almost the highest performance (Figure~\ref{fig:perf_comparision}) once the matrix satisfies the limitations of it. So we arrange DIA ruleset in the first place to pursue the high performance and save time as well. Then it is ELL format that have regular behaviors and relatively easy to predict (shown in Section~\ref{sec:observations}). Naturally, ELL ruleset takes the second place in order in the format search module. Due to COO neither has prominent performance than CSR nor behaves more regular than CSR, there is no reason to place COO ruleset ahead of CSR. Since the parameters related to CSR ruleset have been generated from the feature extraction process of DIA and ELL, so CSR takes the third place, and COO is the last format to predict in the format search module.

In order to unleash the computational power of multi-core architectures, we extend the SMAT system to a multi-threaded version. We employ a coarse parallelism strategy based on task division, which is realized by dividing the input sparse matrix according to the number of threads. To make each thread load-balance, instead of allocating each thread the equal number of rows, we divide the nonzero elements according to the number of threads. In this way, each thread contains almost equal nonzeros, which means the computation operations are almost the same. From~\cite{Zhang_2009}, multi-threaded SpMV achieves better performance in this non-zero scheduling strategy than the scheduling strategies of OpenMP in most cases. In Section~\ref{sec:exp}, the experiments show that our SMAT system also benefits from this parallelism strategy.

According to the nonzeros allocated to each thread, we round up/down the number of nonzeros to make each thread process multiples of rows. Then the input matrix is divided by row, and each thread processes a segment of matrix A, a segment of Y, and the whole X vector. During to the shared memory architecture, since there is no memory conflict, all the three arrays need not to copy but use the initial data. According to this parallelism strategy, SMAT is able to choose different formats among different threads and to achieve the highest performance by combining the best performance of each matrix segment. It is good for the sparse matrices which have non-uniform distribution of the nonzero elements, because these matrices may have different characteristics among the matrix segments.

\section{Experiments and Analysis}
\label{sec:exp}

\subsection{Experimental platforms and data set}
We choose Intel and AMD platforms to test SpMV performance and do analysis. The Intel platform configures Xeon X5680 with 12 cores, one of which have a frequency of 3.33 GHz. The AMD platform configures Opteron 6168 with 24 cores and its frequency is 1.9 GHz.

As for the data set, we use 2373 matrices of UF sparse matrix collection (excluding the ones with complex values and the size of which is smaller than 100) as our data set. The matrix set is divided into training set (2055) and testing set (318), where the testing set is extracted from most matrix groups of the collection in proportion. The training set is used in the offline stage and in the runtime stage we test SMAT performance use the testing set. In the next sections, the performance of SMAT and its accuracy will be illustrated.

\subsection{Performance}
\begin{figure}[htbp]
\begin{tabular}{c|c}
  \includegraphics[scale=0.27]{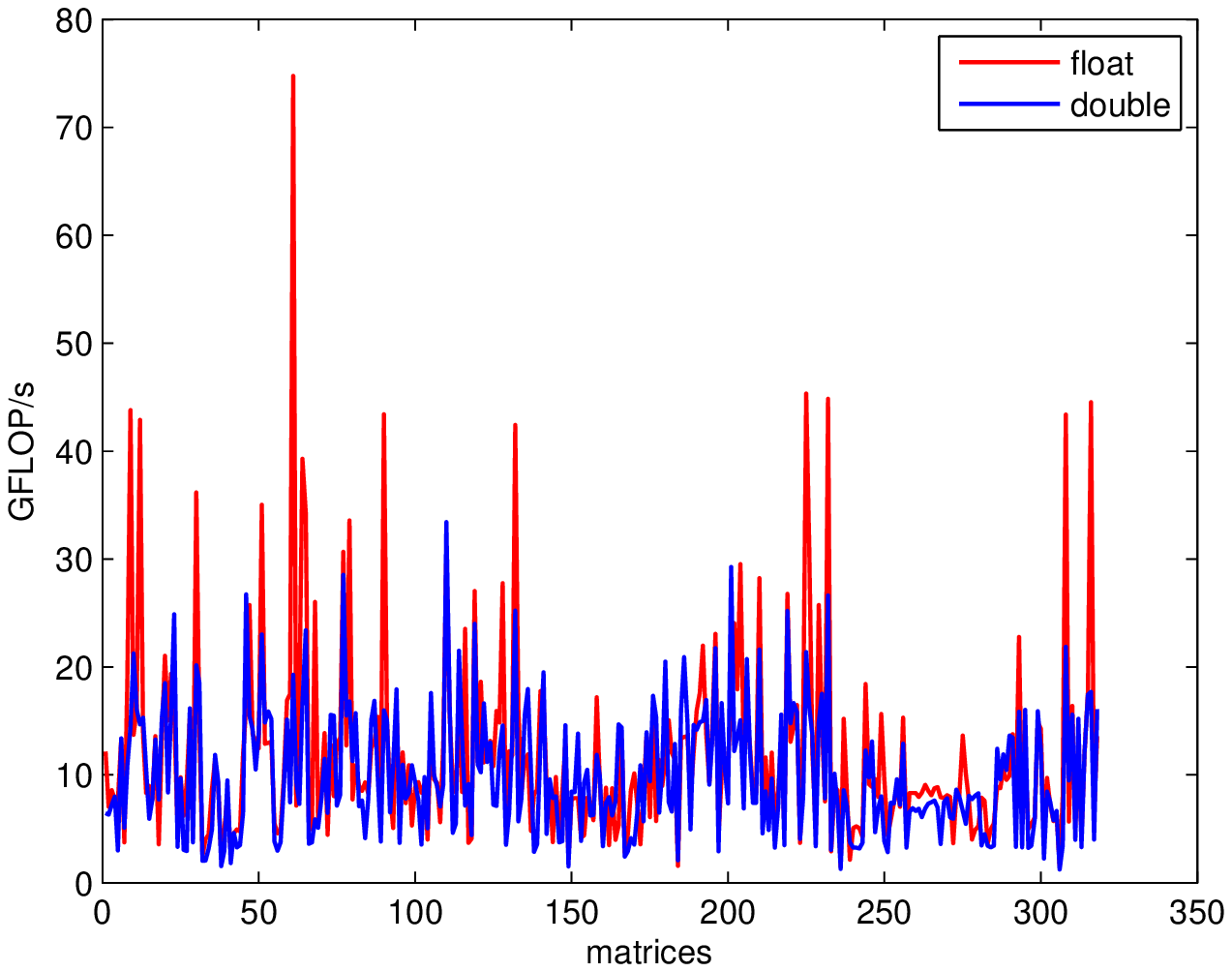}
&
  \includegraphics[scale=0.27]{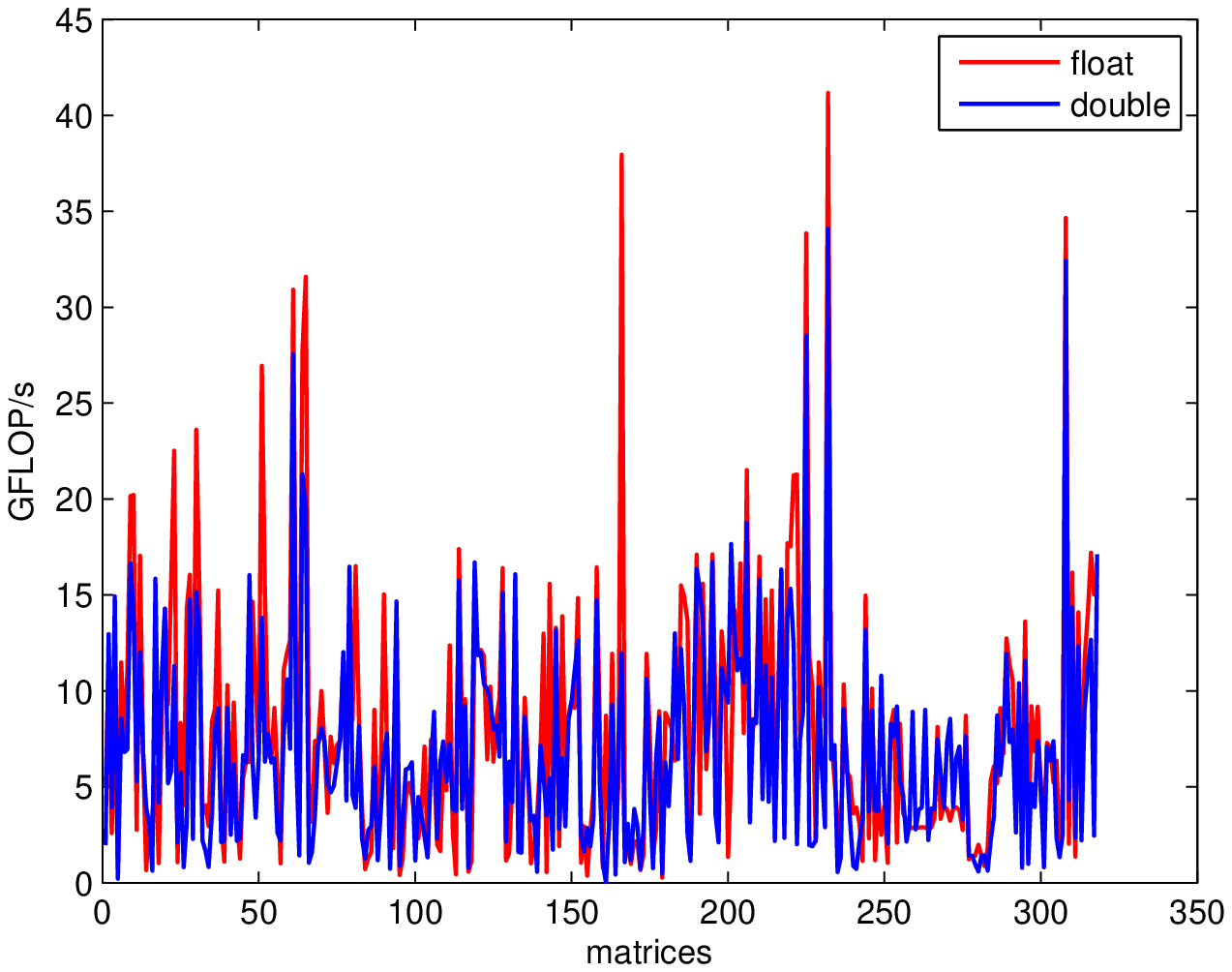}
\\
(a) \small {Intel} & (b) \small {AMD}
  \end{tabular}
  \caption{\small {The highest performance of SMAT on Intel and AMD}}
  \label{fig:best_perf}
\end{figure}

We test the performance of SMAT on both Intel and AMD platforms on the testing sparse matrix set with multiple threads. The best performance of SMAT on Intel and AMD platforms is given in Figure~\ref{fig:best_perf} both with 12 threads. In the two figures, X-axis represents the testing sparse matrices we used (318 matrices), and the Y-axis shows the performance in GFLOP/s. The best performance of SMAT is 75 GFLOP/s with efficiency of 47.3\% in single-precision and 33 GFLOP/s (20.8\%) in double-precision on Intel platform. On AMD platform, SMAT achieves 41 GFLOP/s (22.5\%) in single-precision and 34 GFLOP/s (18.6\%) in double-precision on AMD platform. The peak performance is even higher than the reported maximal performance on GPU (about 18 GFLOP/s)~\cite{Bell:SpMV:NVIDIA:2008}.

According to the figures, both of them show a big variance of the SMAT performance on different matrices due to the diverse features of them. In Figure~\ref{fig:best_perf}(a), the best performance are both obtained when the two matrices (GHS\_indef/linverse and HB/nos7) are stored in DIA format. The same phenomenon is also occurred on AMD platform. This proves DIA-SpMV can achieve good performance provided that the features of the matrix satisfies its limitations.

\begin{figure}[htbp]
\vspace{-0.25cm}
\begin{tabular}{c|c}
  \includegraphics[scale=0.27]{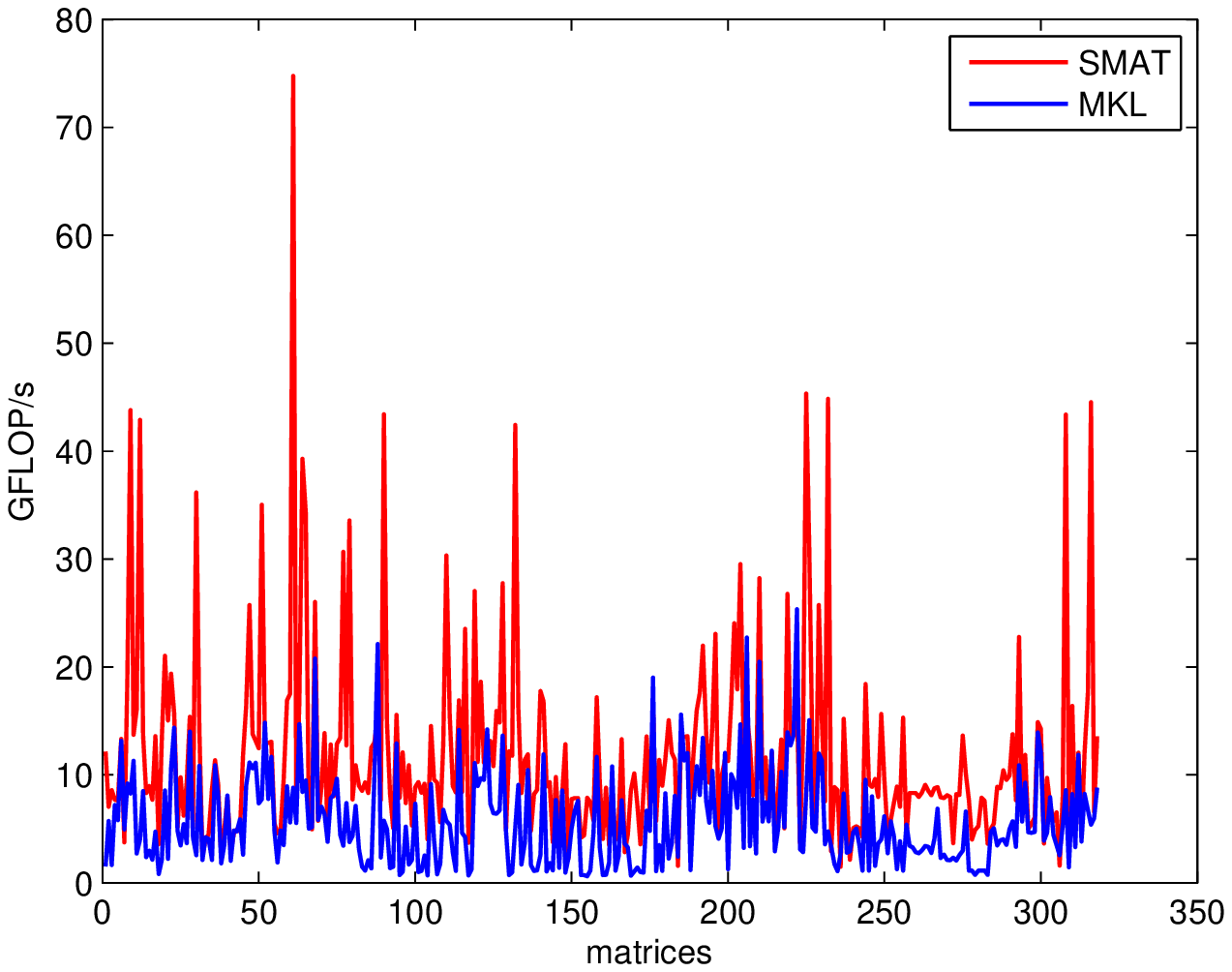}
&
  \includegraphics[scale=0.27]{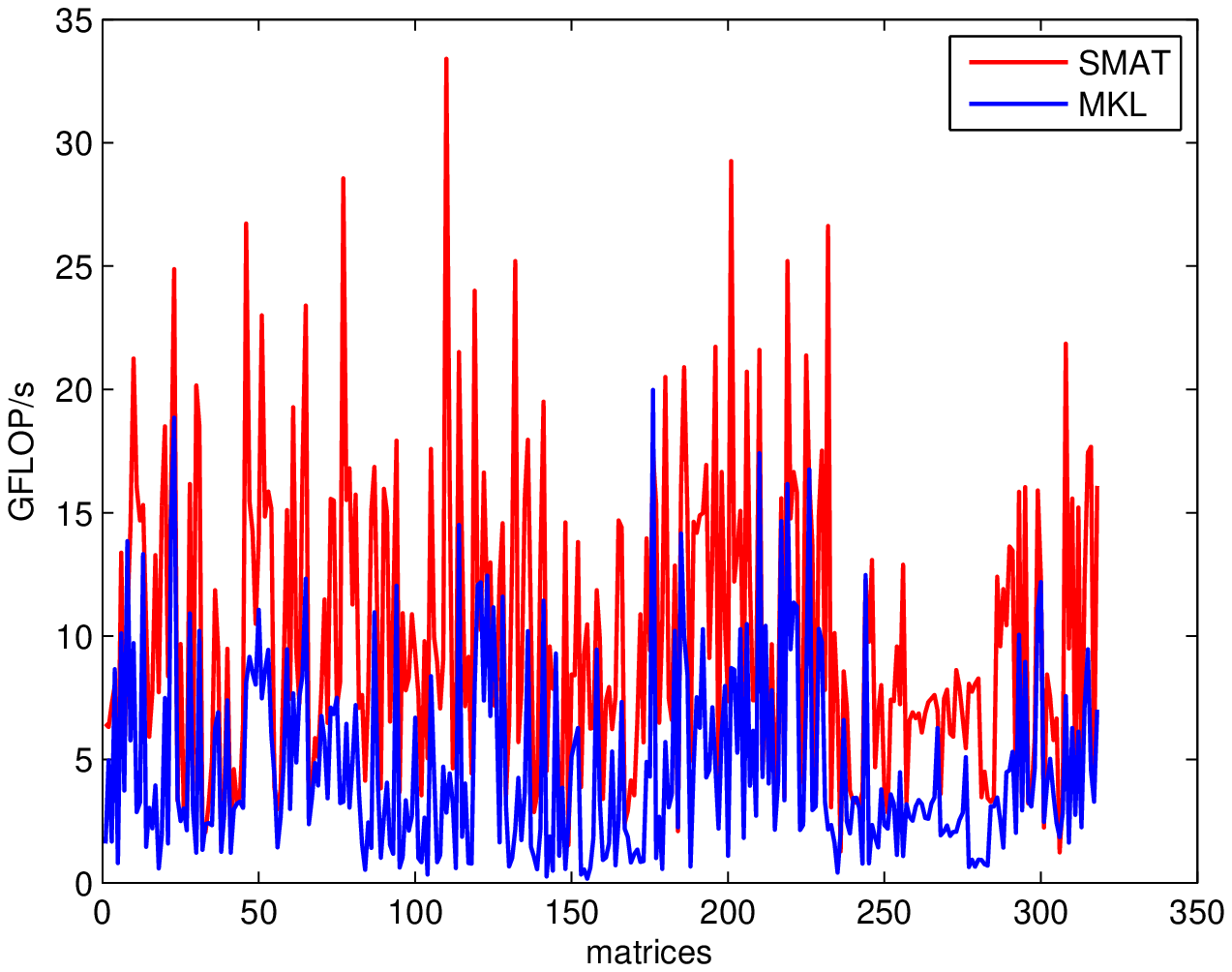}
\\
(a) \small {float} & (b) \small {double}
  \end{tabular}
  \caption{\small {The performance of SMAT and MKL in single- and double-precision on Intel platform}}
  \label{fig:intel_perf_comp}
  \vspace{-0.25cm}
\end{figure}

Considering MKL consists three basic formats (DIA, CSR, and COO), while OSKI~\cite{OSKI} only consists a single CSR format of the four basic formats. We compare the performance of SMAT with MKL library~\cite{MKL} on Intel platform. For justice, we use DIA, CSR, and COO formats of MKL as references, and choose the best performance from them. SMAT and MKL are both executed with 12 threads, which is considered behaving the best performance in most of the time according to our experiments. Their performance in single- and double-precision is shown in Figure~\ref{fig:intel_perf_comp}. From this figure, the average speedup of SMAT to MKL is 3.2 times in single-precision and 3.8 in double-precision. Therefore, it is necessary to apply the SMAT system to application to automatically generate the best SpMV format and implementation and achieve the competitive performance. Though in most of time SMAT obtains better performance than MKL, however, there are still a few of matrices achieve higher performance using MKL, such as matrix Pajek/IMDB.

\subsection{Analysis}
After given the performance advantage of the SMAT system, we analyze the usability of it in two aspects. We first analyze the accuracy of SMAT, and then the prediction overhead of it is evaluated.

\subsubsection{Accuracy analysis}
\begin{figure}[htbp]
\vspace{-0.25cm}
\centering
  \includegraphics[scale=0.35]{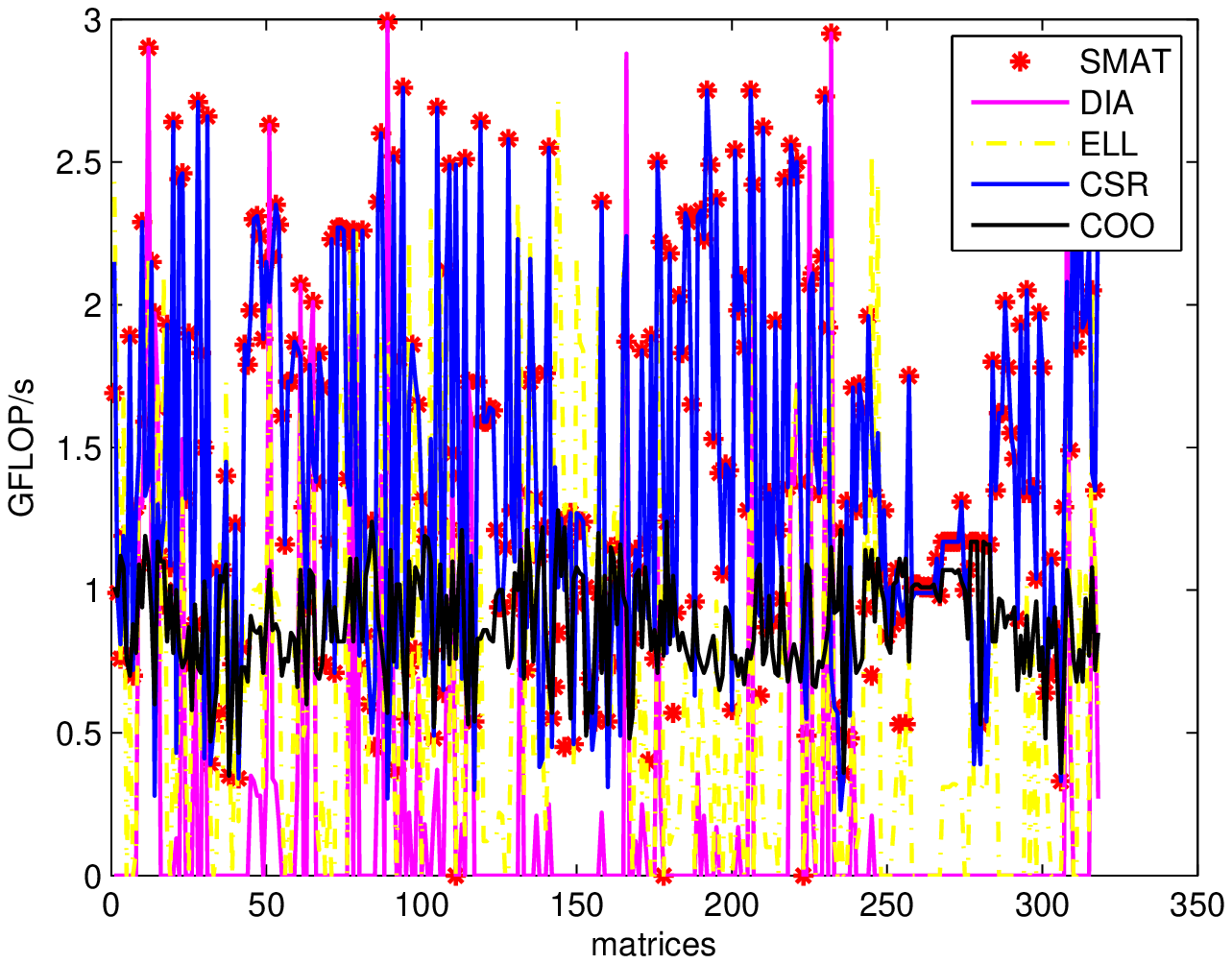}
  \caption{\small {Performance comparision between SMAT and the four formats}}
  \label{fig:accuracy_perf}
  \vspace{-0.25cm}
\end{figure}

We analyze the accuracy of SMAT in two criterions: the estimated format and the achieved performance. The accuracy of SMAT in single- and double-prediction on Intel and AMD platforms is given in Table~\ref{tab:accuracy}, which is 82\%-92\%. From the numbers, it seems SMAT is not reliable enough. However, as we know, when the estimated format of SMAT is the same with the actual best format, the estimate is called accurate. There are still some situations that more than one format achieve similar SpMV performance (the performance difference is smaller than 0.1 GFLOP/s). That means the format accuracy as a criterion of the accuracy of SMAT is neither the unique not the decisive factor. Therefore, we measure the accuracy of SMAT using its performance. The performance of serial SMAT on Intel is shown in Figure~\ref{fig:accuracy_perf} with the performance of the four formats we used in SMAT. The figure shows SMAT achieves the highest performance in 95\% and 90\% matrices in single- and double-precision on Intel, which is higher than the format accuracy of SMAT in Table~\ref{tab:accuracy}. Thus, even though the estimated format of SMAT is not the actual best format, SMAT can also achieve the similar performance. Due to the good accuracy of SMAT, it is reliable to be applied in numerical solvers and realistic applications.

\begin{table}[htbp]
\centering
{\scriptsize
\caption{\small {The accuracy of SMAT on Intel and AMD}}
\label{tab:accuracy}
\begin{tabular}{|c|c|c|c|c|}
\hline
{\bf } & Intel float & Intel double & AMD float & AMD double \\\hline
{\bf SMAT accuracy} & 92\% & 82\% & 85\% & 82\% \\\hline
\end{tabular}
}
\vspace{-0.25cm}
\end{table}

\subsubsection{Prediction overhead}
In the runtime stage of SMAT, we predict the combination of the best format and implementation, which takes extra time compared with pure SpMV implementation. Thus, the prediction overhead of SMAT should be decreased as much as possible. The prediction overhead is tested and measured as the times of CSR-SpMV with simple implementation. The numbers are given in Table~\ref{tab:times} along with the times of ``brute-force" search, in which all the SpMV implementations of the four formats are executed to choose the highest performance. Compared with ``brute-force" search, SMAT performs significant advantage. From the table, the prediction overhead is 8-19 times of CSR-SpMV.

\begin{table}[htbp]
\centering
{\scriptsize
\caption{\small {Average time overhead of format search in runtime stage (represented by the times of CSR-SpMV)}}
\label{tab:times}
\begin{tabular}{|c|c|c|c|c|}
\hline
{\bf } & Intel float & Intel double & AMD float & AMD double \\\hline
{\bf SMAT prediction} & 17 & 8 & 18 & 19 \\\hline
{\bf brute-force search} & 1868 & 2075 & 2404 & 2522 \\\hline
\end{tabular}
}
\end{table}

Due to the prediction overhead of SMAT, an application will benefit from SMAT when performing the SpMV kernel with a single sparse matrix repeatedly. Now we estimate the limitation times of SpMV called in the realistic applications. We use $t_{CSR}$ to represent the execution time of a CSR-SpMV, and $t_{SMAT}$ represents the execution time of the best implementation generated by SMAT. Assume $n$ times of SpMV with a single matrix will be taken in an application. We compare the execution time before and after using SMAT to calculate the limitation number. The relation is shown in Equation~\ref{equ:num_iter}, where 19 is the maximum prediction overhead represented in times of CSR-SpMV in Table~\ref{tab:times}, and 3 is the minimum speedup of SMAT to MKL. From this equation, the number of calls should be larger than 29 that an application can benefit its performance from SMAT. Considering there are plenty of algorithms, like iterative algorithms, need to call SpMV kernel for hundreds of times~\cite{Kelley_1995}. The prediction overhead of the SMAT system is acceptable and applicable to these numerical solvers and applications.

\vspace{-0.4cm}
\begin{equation}\label{equ:num_iter}
\footnotesize
    19 * t_{CSR} + (n/3) *t_{SMAT} < n * t_{CSR} \Longrightarrow n > 29
\end{equation}

\section{Related Work}
\label{sec:related_work}

Plenty of work has already implemented to optimize SpMV and improve its performance. The optimizations fall into two categories, one is applying new storage formats, the other is applying architecture specific optimizations~\cite{clspmv}. The storage format optimization will decrease the memory accesses and improve SpMV performance. Eun-jin Im et al.~\cite{sparsity} created BCSR format to better develop the performance of dense blocks in a sparse matrix. Richard Vuduc et al.~\cite{OSKI} improve BCSR format to VBR to better develop the performance of dense blocks with different sizes. Kourtis et al.~\cite{CSX} proposed CSX as a generalized approach to compress metadata by exploiting substructures within the matrix. In the other aspect, some people optimize SpMV considering hardware characteristics, especially the novel multi-core CPU, FPGA, and many-core GPU. Nagar et.al.~\cite{Krishna_2011} customized SpMV on Convey HC-1 which is a self-contained heterogeneous system combined a Xeon-based host and an FPGA-based co-processor. Williams et al. ~\cite{Williams_2009} evaluated different optimization strategies on five platforms (AMD Opteron, Intel Clovertown, Sun Niagara2, and STI Cell SPE). Bell and Garland~\cite{Bell:SpMV:NVIDIA:2008} optimized different SpMV kernels with different sparse matrix formats on NVIDIA GPUs, and also presented a new format (HYB). All of these optimizations contribute to improve SpMV performance.

Among the optimization work of SpMV, there are also lots of work use auto-tuning method to increase its performance and portability among different architectures. Richard Vuduc et al.~\cite{OSKI} built an auto-tuner named OSKI to tune the block size for a matrix in BCSR or VBR formats. Williams et al.~\cite{Williams_2009} use auto-tuning method with a hierarchy strategy to choose the best parameter combinations. Choi et al.~\cite{Choi_2010} implemented Blocked Compress Sparse Row (BCSR) and Sliced Blocked ELLPACK (SBELL) formats on Nvidia GPUs, and tuned the block size of them. Xintian Yang et al.~\cite{Yang_2011} proposed a mixed format and automatically chose the partition size of each format with the help of a model. The work above all utilized auto-tuning method to tune a unique matrix format through evaluating different sizes considered architecture characteristics.

Bor-Yiing Su et al.~\cite{clspmv} proposed a hybrid format (Cocktail) to split a matrix and took advantage of the strengths of many different sparse matrix formats. Though clSpMV also tunes for the best composition of the Cocktail format, they valued each possible format considered the typicality of the formats instead of the architecture features. The architecture features related to each format have already been considered in the off-line stage of clSpMV. This method is similar to the SMAT system. However, there are still some crucial differences between them. First and the most important, in the online decision making stage, clSpMV uses the maximum GFLOPS measured in offline stage under the current settings. In some situations, the best performance of one format cannot represent the performance of all the matrices in this format, because different matrix features have a huge influence on the performance. It is more accurate to use the features of each input matrix to predict its best format rather than a single maximum performance of each format. Second, we use realistic matrices from UF collection as the training data to generate decision tree through data mining method, which is more reliable than hand-generated decision making process. Last, we extract more matrix features than clSpMV, while it utilizes more matrix formats. In the future, we'll extend the SMAT system with more formats.

\section{Conclusion}
\label{sec:conclusion}

In this paper, we propose an SpMV auto-tuner (SMAT) to generate a high performance SpMV library that can be easily adopted by numeric solvers and applications. The previous optimizations or auto-tuning libraries provide various storage format interfaces to users that seems to be flexible. However, it is the ``flexibility" that hinders their popularity. With respect to the fact that most of existing programs are developed based on compressed sparse row (CSR) formats, SMAT provides an unified interface based on CSR to programmers. It choose the best format and the fastest implementation of any input sparse matrix in runtime. The mechanism behind is a data-mining model that is built on a set of performance critical parameters for sparse matrices. The experiments in multi-core X86 processors show that SMAT achieves impressive performance on both Intel and AMD multi-core platforms. The achieved peak performance is more than 30 GFLOP/s in double-precision, that is even higher than the reported maximal performance on GPU (about 18 GFLOP/s)~\cite{Bell:SpMV:NVIDIA:2008}. Although the on-line search overhead is about $8-19$ times of execution time of CSR-SpMV, it can be amortized in numerical solvers that often have hundreds of SpMV calls. In the future, we will add more matrix formats to the SMAT system and integrate it to realistic applications.






\bibliographystyle{abbrvnat}


\softraggedright

\raggedright
\scriptsize
\bibliography{SMAT_JiajiaLi}



\end{document}